\RequirePackage{amsmath}
\documentclass{svproc}
\usepackage{url}
\usepackage{graphicx}
\usepackage{amsmath}

\newcommand{\msun}{\ensuremath{M_\odot}}

\newcommand{\nue}{\ensuremath{\nu_{e}}}
\newcommand{\nuebar}{\ensuremath{\bar \nu_e}}
\newcommand{\numt}{\ensuremath{\nu_{\mu\tau}}}
\newcommand{\numtbar}{\ensuremath{\bar \nu_{\mu\tau}}}
\newcommand{\numu}{\ensuremath{\nu_{\mu}}}
\newcommand{\nutau}{\ensuremath{\nu_{\tau}}}
\newcommand{\numubar}{\ensuremath{\bar \nu_{\mu}}}
\newcommand{\nutaubar}{\ensuremath{\bar \nu_{\tau}}}
\newcommand{\mev}{\mbox{MeV}}

\newcommand{\gcc}{\ensuremath{{\mbox{g~cm}}^{-3}}}

\newcommand{\Chimera}{{\sc Chimera}}

\begin{document}
\mainmatter              
\title{On the character of turbulence in self-consistent models of core-collapse supernovae}
\titlerunning{Turbulence in core-collapse supernovae}  
%
\author{Jordi Casanova\inst{1}, Eirik Endeve\inst{4,2,3,7}, Eric J. Lentz\inst{2,1,3}, O. E. Bronson Messer\inst{5,1,2}, W. Raphael Hix\inst{1,2}, J. Austin Harris\inst{1,5}, \and Stephen W. Bruenn\inst{6}}
%
\authorrunning{Jordi Casanova et al.} 
%
\tocauthor{Ivar Ekeland, Roger Temam, Jeffrey Dean, David Grove,
Craig Chambers, Kim B. Bruce, and Elisa Bertino}
\institute{Physics Division, Oak Ridge National Laboratory, P.O. Box 2008, Oak Ridge, TN 37831-6354, USA\\
\email{jordi.casanova@upc.edu},\\ 
\and
Department of Physics and Astronomy, University of Tennessee, Knoxville, TN 37996-1200, USA\\
\and
Joint Institute for Computational Sciences, Oak Ridge National Laboratory, P.O. Box 2008, Oak Ridge, TN 37831-6173, USA\\
\and
Computer Science and Mathematics Division, Oak Ridge National Laboratory, P.O. Box 2008, Oak Ridge, TN 37831-6164, USA\\
\and
National Center for Computational Sciences, Oak Ridge National Laboratory, P.O. Box 2008, Oak Ridge, TN 37831-6164, USA\\
\and
Department of Physics, Florida Atlantic University, 777 Glades Road, Boca Raton, FL 33431-0991, USA\\
\and
This manuscript has been authored in part by UT-Battelle, LLC, under contract DE-AC05-00OR22725 with the US Department of Energy (DOE). The US government retains and the publisher, by accepting the article for publication, acknowledges that the US government retains a nonexclusive, paid-up, irrevocable, worldwide license to publish or reproduce the published form of this manuscript, or allow others to do so, for US government purposes. DOE will provide public access to these results of federally sponsored research in accordance with the DOE Public Access Plan (http://energy.gov/downloads/doe-public-access-plan).}

\maketitle              

\begin{abstract}
	Neutrino-driven convection plays a crucial role in the development of core-collapse supernova (CCSN) explosions. However, the complex mechanism that triggers the shock revival and the subsequent explosion has remained inscrutable for many decades. Multidimensional simulations suggest that the growth of fluid instabilities and the development of turbulent convection will determine the morphology of the explosion. We have performed 3D simulations using spherical-polar coordinates covering a reduced angular extent (90$^\circ$ computational domain), and with angular resolutions of 2$^\circ$, 1$^\circ$, 1/2$^\circ$, and 1/4$^\circ$, to study the development of turbulence in core-collapse supernova explosions on a time scale of order $100$~ms. We have employed the multi-physics \Chimera\ code that includes detailed nuclear physics and spectral neutrino transport. Coarse resolution models do not develop an inertial range, presumably due to the bottleneck effect, such that the energy is prevented from cascading down to small scales and tends to accumulate at large scales. High-resolution models instead, start to recover the $k^{-5/3}$ scaling of Kolmogorov's theory. 
Stochasticity and few simulation samples limit our ability to predict the development of explosions. 
Over the simulated time period, our models show no clear trend in improving (or diminishing) conditions for explosion as the angular resolution is increased.  
However, we find that turbulence provides an effective pressure behind the shock ($\sim$40--50\% of the thermal pressure), which can contribute to the shock revival and be conducive for the development of the explosion. Finally, we show that the turbulent energy power spectrum of reduced angular extent and full 4$\pi$ models are consistent, thus indicating that a 90$^\circ$ computational domain is an adequate configuration to study the character of turbulence in CCSNe. 
\keywords{turbulence --- neutrinos --- nucleosynthesis --- stars: evolution --- stars: massive  --- supernovae: general}
\end{abstract}
\section{Introduction}

Massive stars ($M > 8 M_{\odot}$) explode as core-collapse supernovae (CCSNe) and release kinetic energies of $\sim$ 1~B (1~B = $10^{44}$~J) \cite{Smar09}. The iron core collapses and matter bounces back once nuclear densities are achieved ($\rho>10^{14}$~\gcc). The shock wave eventually stalls after tens of milliseconds, due to photodissociation and escaping neutrinos. The delayed-neutrino mechanism is believed to play an important role, as neutrinos heat the material in the gain (or net neutrino heating) region, revive the stalled shock and power the final explosion. Therefore, in the neutrino-driven convection context, neutrinos are decisive in powering the explosion by depositing a small fraction of their energy in the gain region. Owing to neutrino heating, the base of the gain region becomes unstable to convection, which is fully developed within $\sim$ 50-100~ms.  The growth of fluid instabilities and the development of convection enhances the effective neutrino heating by increasing the dwell time of a fluid element in the heating region. The morphology of the flow depends on the numerical resolution employed, with low-resolution models exhibiting a few, long-lived large plumes that occupy the entire gain region and high-resolution models presenting smaller structures that are short-lived. 

Detailed spherically symmetric simulations do not produce explosions \cite{RaJa00,LiMeTh01,LiRaJa05}.  
Spherically symmetric simulations with parameterized neutrino transport can be used to reproduce explosion properties and nucleosynthetic yields, mainly by adding an extra heating source into the system: piston \cite{WoWe95}, thermal bomb \cite{ThNoHa96}, absorption methods \cite{FrHaLi06}, light-bulb based schemes \cite{UgJaMa12}, or the PUSH method \cite{PeHeFr15}.  Nonetheless, simulations in spherical symmetry exclude multi-D effects, such as fluid circulation and instabilities, that are critical to investigate the nature of the explosion. Recent multi-D studies have revealed several processes that result in turbulent flows and may aid in reviving the shock, such as the standing accretion shock instability (SASI; \cite{BlMeDe03,HaMaMu12}), neutrino-driven convection in the gain region \cite{HeBeCo92,HeBeHi94,Jank12,KoTaSu12,HaMaMu12}, and progenitor asphericities \cite{CoOt13,MuJaMa12}. Furthermore, it is believed that turbulence contributes to the pressure that counteracts the ram pressure of the infalling material, thus aiding power the final explosion. Previous multi-D studies have shown that the turbulent pressure is $\sim$30-50$\%$ of the total thermal pressure \cite{CoOt15,MuDoBu13,RaOtAb16}. 

Radice et al. \cite{RaOtAb16} explored the role of turbulence in the context of neutrino-driven convection. They employed a high-order finite-difference scheme, as implemented in the WhiskyTHC code \cite{RaRe12,RaReGa14}. Their simulations were performed with simplified physics, such as a fixed accretion rate at the shock, point-mass gravity that approximates the influence of the proto-NS, and parameterized neutrino heating and cooling. The study presented 3D simulations in spherical-polar coordinates using a reduced angular extent (90$^\circ$ wedge format) with constant angular resolutions of 1.8$^\circ$, 0.9$^\circ$, 0.45$^\circ$, 0.3$^\circ$, 0.15$^\circ$, and 0.09$^\circ$. They found that low-resolution calculations truncated the inertial range, which affected interactions between eddies of different scales and interrupted the energy transfer down to small scales \cite{ShJa93,FrKuPa08,DoHaYo03}. These low-resolution models presumably suffer from the bottleneck effect, which tends to accumulate energy at large-scales.
At high resolutions, the turbulent energy power spectrum presents a region compatible with Kolmogorov's $k^{-5/3}$ scaling (especially for the model with an angular resolution of 0.09$^\circ$). 
Moreover, turbulence contributes with an extra pressure ($\sim$ 30-40\% of the thermal pressure), that may help revive the shock.

We report on self-consistent simulations of CCSNe, initiated from a 15~\msun\ progenitor and evolved fully convective for about $150$~ms, in 3D with angular resolutions of $2^\circ$, $1^\circ$, $1/2^\circ$, and $1/4^\circ$ using our neutrino radiation hydrodynamics ($\nu$RHD) code \Chimera\ \cite{BrBlHi18} to study the evolution and possible impacts of turbulence in high-fidelity CCSN models.  As the work presented by \cite{RaOtAb16} employed simplified physics, a reanalysis of the effect of turbulence as function of the numerical resolution in the context of neutrino-driven convection is warranted. We employed the same reduced angular extent (90$^\circ$ wedge format) utilized in \cite{RaOtAb16}, to reach high spatial resolution and limit computational cost. Our results are consonant with the work of \cite{RaOtAb16}, and our findings confirm that high resolution is necessary to resolve the inertial range.  Our most refined models (e.g., $\delta\theta$=$\delta\phi$=1/2$^\circ$) start recovering the $k^{-5/3}$ scaling of Kolmogorov's theory. Furthermore, we also find that turbulence contributes additional pressure ($\sim$ 40-50\% of the thermal pressure) behind the shock, which could contribute to the shock revival. Thus, resolving turbulence is crucial for understanding the fate of the models.  
We also show a comparison between a full 4$\pi$ and a 90$^\circ$ wedge model using the same angular resolution of 2$^\circ$.  
This comparison indicates that a 90$^\circ$ computational domain sufficiently captures the flow on the smallest resolved scales.  
The results presented here may help inform resolution requirements for planned CCSN models.  

\section{Input physics and initial setup}

We employed our neutrino radiation hydrodynamic code \Chimera, a multi-physics, parallel, code built specifically for multidimensional simulation of CCSNe in spherical-polar coordinates ($r,\theta,\phi$) \cite{BrBlHi18}.
Hydrodynamics is evolved via a dimensionally split, Lagrangian-plus-remap Newtonian scheme with piecewise parabolic reconstruction (PPMLR \cite{CoWo84}) as implemented in VH1 \cite{HaBlLi12}. 
Self-gravity is computed in spherical symmetry with a general relativistic (GR) potential for the wedge models with the $4\pi$-model enhanced by a Newtonain multipole expansion \cite{MuSt95,MaDiJa06}. 
The neutrino transport solver is an improved and updated version of the multi-group (frequency) flux-limited diffusion (MGFLD) transport solver of Bruenn \cite{Brue85}, enhanced for GR \cite{BrDeMe01}.

\begin{table*}
\caption{Main characteristics of the reduced angular extent models (series-D) and full model (series C).}
\label{table:1}
\centering
\begin{tabular}{c c c c c }
\hline\hline

Model & $\delta\theta$=$\delta\phi$ ($^\circ$) & t$_{\rm initial}$ (ms) & t$_{\rm final}$ (ms) & $\delta_{\rm FFT}$ (km)\\
\hline
 D15-W2.00-3D & 2.00 & 100 & 285.0 & 2.5$\sqrt{2}$\\
 D15-W1.00-3D & 1.00 & 100 & 297.0 & 1.25$\sqrt{2}$\\
 D15-W0.50-3D & 0.50 & 100 & 282.0 & 0.625$\sqrt{2}$\\
 D15-W0.25-3D & 0.25 & 150 & 195.0 & 0.3125$\sqrt{2}$\\
 C15-3D-2deg & 2.00 & 1.3 & 440.0 & 2.5$\sqrt{2}$\\
\hline
\end{tabular}
\end{table*}

We solve for all three flavors of neutrinos and anti-neutrinos with four coupled species: \nue, \nuebar, $\numt=\{\numu,\nutau\}$, $\numtbar=\{\numubar,\nutaubar\}$, using 20 logarithmically spaced energy groups $\alpha\epsilon = 4$--250~\mev, where $\alpha$ is the lapse function and $\epsilon$  the comoving-frame group-center energy.
The neutrino--matter interactions used are the full set of \cite{BrLeHi16}.
We utilize the LS220 EoS \cite{LaSw91} (incompressibility $K = 220$~\mev) for  $\rho>10^{11}$~\gcc\ and an enhanced version of the Cooperstein EoS \cite{Coop85} for $\rho<10^{11}$~\gcc.
In regions out of NSE, we use an alpha network, consisting of 17 nuclear species \cite{HiTh99b} designed to capture the important isotopes present in supernova conditions.

Simulations employ a 15~\msun\ pre-supernova progenitor \cite{WoHe07}, initialized from a 1D simulation at t=100~ms post-bounce by introducing a density perturbation interior to the shock, mimicking perturbations seen in multi-D simulations evolved from bounce. We performed spherical-polar simulations in a reduced angular extent that covered $\theta\in[\pi/4,3\pi/4]$ and $\phi\in[0,\pi/4]$. We employed different constant angular resolutions (i.e., $\delta$$\theta$=$\delta$$\phi$): $360^{2}$ uniform zones ($\delta\theta$=$\delta\phi$=1/4$^\circ$) for D15-W0.25-3D, $180^{2}$ uniform zones ($\delta\theta$=$\delta\phi$=1/2$^\circ$) for D15-W0.50-3D, $90^{2}$ uniform zones ($\delta\theta$=$\delta\phi$=1$^\circ$) for D15-W1.00-3D, and $45^{2}$ uniform zones ($\delta\theta$=$\delta\phi$=2$^\circ$) for D15-W2.00-3D. In Table 1, we describe the properties of each model: name of the model, angular resolution employed ($\delta\theta$=$\delta\phi$), initial time at which we start the multi-D simulations (t$_{\rm initial}$), final time of the simulations (t$_{\rm final}$), and grid spacing on the Cartesian box employed to perform the Fourier analysis ($\delta_{\rm FFT}$). The inner region (10,700 km; 2.32~\msun) is remapped onto 540 radial shells, except for D15-W0.25-3D that used 1080 radial zones instead, on logarithmic radial grid ($\delta r/r$) modified to track density gradients for all the models. C15-3D-2deg is a full 4$\pi$ model from Series C (see Section 4.3 for more details). During the evolution, the radial zones are gradually and automatically repositioned during the remap step to follow changes in the radial structure.

\section{Simulation overwiew}

\begin{figure}[t!]
    \centering
    \includegraphics[width=0.9\textwidth]{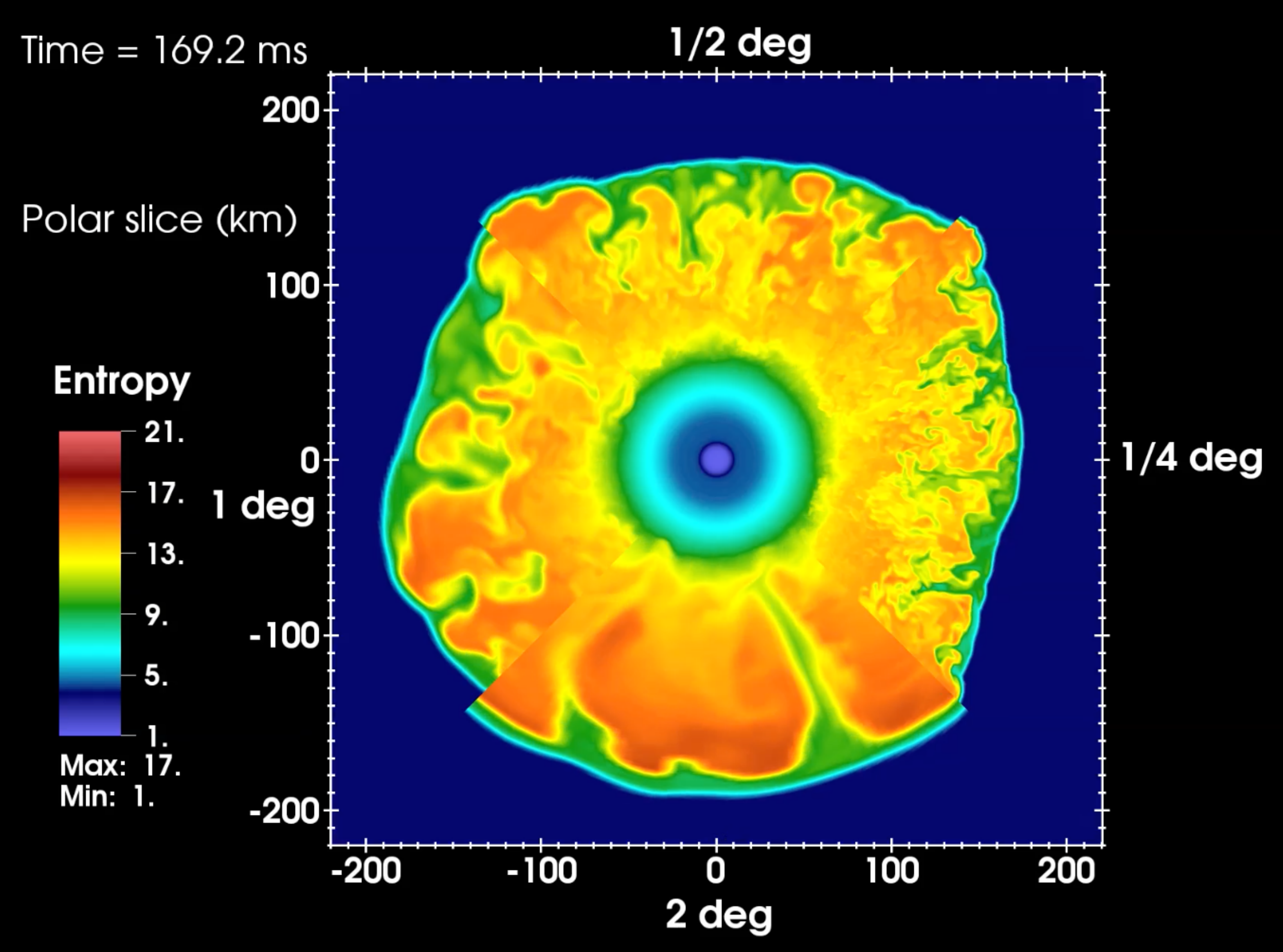}
    \caption{Polar slice of the entropy profile for the 3D wedge models at t=169.2~ms. Upper quadrant corresponds to D15-W0.50-3D, left quadrant to D15-W1.00-3D, lower quadrant to D15-W2.00-3D, and right quadrant to D15-W0.25-3D.}
\label{fig:Entropy}
\end{figure}

The 3D simulations were started from a 1D run at t=100~ms, relative to core bounce, except for D15-W0.25-3D, which was started from D15-W0.50-3D at t=150~ms. The gain region is the region above the proto-NS, within the shock cavity, where neutrino heating rates dominate over neutrino cooling rates. The gain region becomes unstable to convection when neutrino heating at the base results in the formation of initial plumes, which grow against the infalling material onto the proto-NS. Rising plumes arise in the gain region at t$\sim$130~ms and begin deforming the shock. 
The shock is deformed from spherical as the initial plumes reach the shock followed by the development of an convective circulation.
Low-resolution models give rise to a few big plumes that occupy the entire gain region, while in high-resolution models, large-scale structures are prevented from forming and plumes break into smaller features, resulting in a more complex flow.
A polar slice of the entropy profile (Figure~\ref{fig:Entropy}) show the development of full convection in the gain region for the wedge models at t=169.2~ms.

\begin{figure}[hp!]
    \centering
    \includegraphics[width=0.9\textwidth,height=9.5cm]{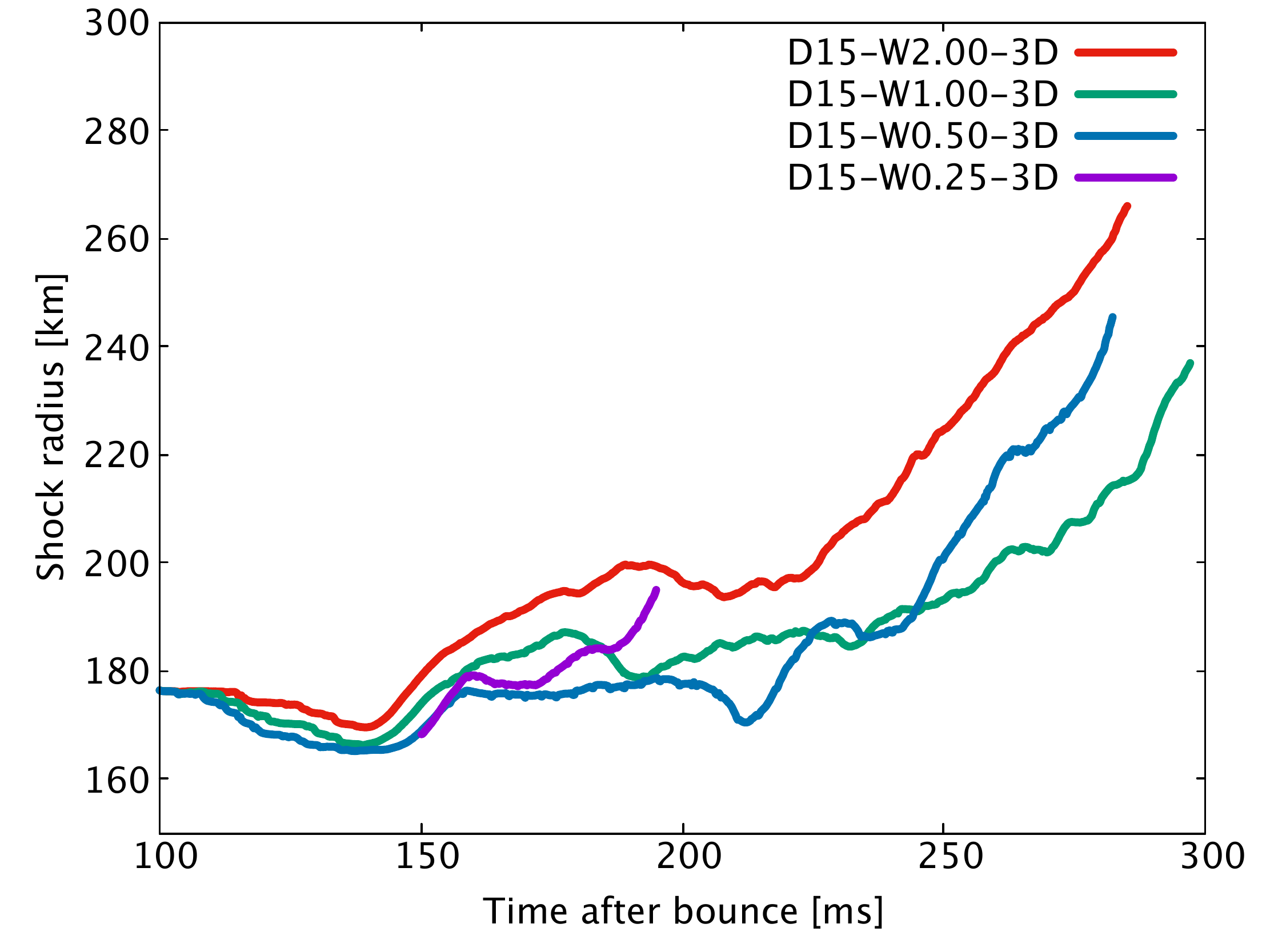}\\
    \includegraphics[width=0.9\textwidth,height=9.5cm]{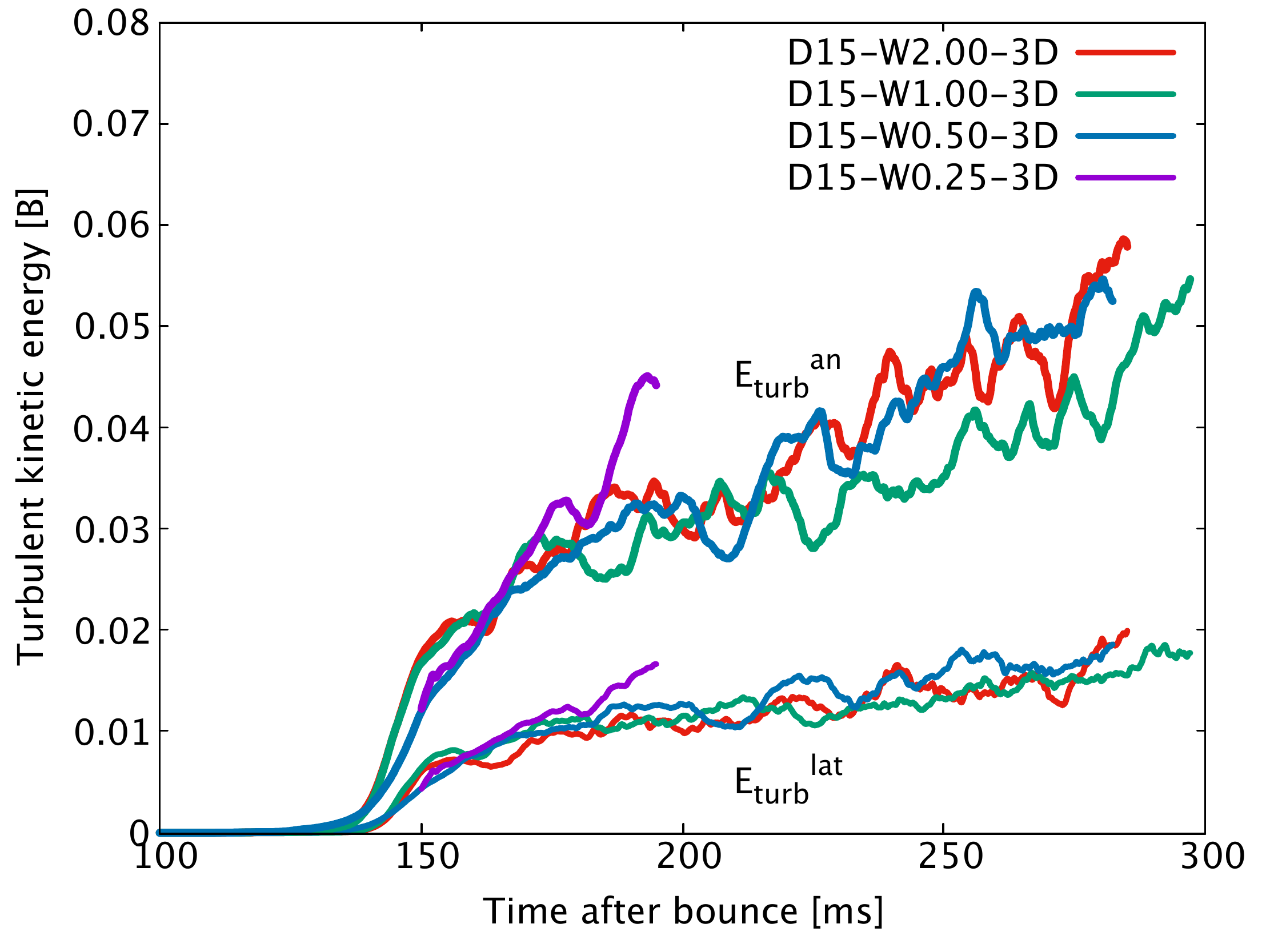}\\
    \caption{Upper panel: average shock radius in function of time after bounce for the 3D models (D15-W2.00-3D, D15-W1.00-3D, D15-W0.50-3D, and D15-0.25-3D). Models were stopped at t$\simeq$300~ms, except for the D15-W0.25-3D that was stopped at 195~ms. Lower panel: anisotropic and lateral turbulent kinetic energy for the 3D models.}
\label{fig:Shock-KineticEn}
\end{figure}

Within 50~ms of initiation, a fully convective state is achieved in the four computed wedge models.
However, we do not see any conclusive trend with resolution in the shock growth during this period (see upper panel of Figure~\ref{fig:Shock-KineticEn} for the evolution of the average shock radii for each model).
After t$\sim$200~ms, the shock radii starts increasing steadily until the end of the simulations. The rising plumes reach the shock and deform it, as secondary Raleigh-Taylor instabilities arise all along the gain region. 
The shock radii for all models start increasing progressively through the end of the simulations.
The shock radius is a good parameter for exploring the explodability of the models, as a more robust extended shock is a potential indicator of an earlier explosion\footnote[1]{One conventional definition of an explosion is when the shock reaches 500~km.}.
D15-W2.00-3D presents a larger mean shock radius increase, possibly indicating the development of an earlier explosion. High-resolution models present weaker extended shocks, though at t$\sim$250~ms, the mean shock radius for D15-W0.50-3D increases faster than that of D15-W1.00-3D, which may signal an earlier explosion. D15-W0.25-3D also shows a vigorous shock evolution, however it is worth noting that this model was initiated at t$\sim$150~ms, a time at which the initial plumes are already developed, and was evolved for only 45~ms due to cost.

In the lower panel of Figure \ref{fig:Shock-KineticEn}, we show the turbulent kinetic energy in the gain region for the 3D models, $E_{\rm turb}^{X}$, which measures the energy associated with the disordered flow:
%
 \begin{align}
 E_{\rm turb}^{X}= \int_{\rm{gain}} \frac{1}{2}\rho\mid{\mathbf{v}^{X}}\mid^{2} \, dV,
 \end{align}
%
where $\mathbf{v}^{X}$ are the velocity components in the gain region. 
The anisotropic turbulent kinetic energy, $E_{\rm turb}^{\rm an}$, includes the contribution of all the velocity components with the $v_{r}$ component measured as $v_{r}^{\rm an} = (v_{r} - \langle v_{r}\rangle)$. The lateral turbulent kinetic energy, $E_{\rm turb}^{\rm lat}$, is calculated by removing the radial velocity component $v_{r}$ in $v^{\rm lat}= (\overline{v}-v_{r}\widehat{e}_{r}$), and thus, it only includes the angular velocity components. Once convection sets in after t$\sim$150~ms, the competition
 between rising plumes and the infalling material results in lateral motions, and therefore, the lateral turbulent kinetic energy starts increasing. The turbulent kinetic energy provides another good diagnostic for the explodability of the models, as larger kinetic energies tend to correlate with the onset of explosions. Presumably, a higher turbulent pressure would contribute to the shock revival \cite{CoOt15}. The turbulent kinetic energy is larger for D15-W2.00-3D, potentially indicating an earlier shock revival, while it is smaller for high resolution models, thus signaling more likely delayed explosions. However, D15-W2.00-3D and D15-W0.50-3D present similar energy profiles. The latter model exhibits a larger turbulent kinetic energy than D15-W1.00-3D, possibly as a consequence of the stochastic behavior induced by turbulence.

\section{Energy cascade}
\subsection{Methodology}

We computed the turbulent kinetic energy power spectrum by Fourier transform of the velocity components as found in \cite{RyJoFr00,EnCaBu12,RaOtAb16}, and using the FFTW library\footnote[2]{http://www.fftw.org}. We mapped the polar-spherical grid onto a Cartesian grid and selected a cubic box of (50$\sqrt{2}$~km)$^{3}$ placed within the gain region,  interpolating the velocity to the analysis box by trilinear interpolation.
We started with a box spacing ($\delta_{\rm FFT}$ in Table 1) of 2.5$\sqrt{2}$~km for D15-W2.00-3D (i.e., angular resolution of 2$^\circ$), which is similar to the spacing utilized in \cite{RaOtAb16}. The spacing for the rest of the simulations is chosen according to the corresponding angular resolutions: 1.25$\sqrt{2}$~km for D15-W1.00-3D (1$^\circ$), 0.625$\sqrt{2}$~km for D15-W0.50-3D (0.5$^\circ$), and 0.3125$\sqrt{2}$~km for D15-W0.25-3D (0.25$^\circ$). 

We applied a windowing function to enforce periodicity within the cube:
\begin{equation}
 W(x,y,z) = w\Bigg[\frac{(x-x_{0})}{(x_{1}-x_{0})}\Bigg]w\Bigg[\frac{(y-y_{0})}{(y_{1}-y_{0})}\Bigg]w\Bigg[\frac{(z-z_{0})}{(z_{1}-z_{0})}\Bigg],~
\end{equation}
where ($x_{0},y_{0},z_{0}$) and ($x_{1},y_{1},z_{1}$) are the vertices of the cube and $x$, $y$, and $z$ are the coordinates within the box. 

We employed an exponential function following \cite{RaOtAb16}:
\begin{equation}
w(\psi) = \left\{
  \begin{array}{lr}
    0 & \mbox{if~}\psi = 0\\
    \exp\left[\frac{-1}{1-\left(\left(\frac{\psi-\Delta}{\Delta}\right)^2+1\right)}\right] & \mbox{if~} 0 < \psi \le \Delta\\
    1  & \mbox{if~}\Delta < \psi \le (1-\Delta)\\
    \exp\left[\frac{-1}{1-\left(\left(\frac{\psi-(1-\Delta)}{\Delta}\right)^2+1\right)}\right] & \mbox{if~} (1-\Delta) < \psi \le 1\\
    0 & \mbox{if~}\psi = 1
 \end{array}
\right.
\end{equation}
The windowing function equals zero at the edges of the cube and it increases (or decreases) exponentially depending on the value of the parameter $\Delta$, which is arbitrarily set to 3 points.
By this procedure, the initial input quantity remains unchanged for most of the points in the center of the cube.

We calculate the Fourier amplitudes of the velocity components multiplied by the above-mentioned windowing function:
\begin{equation}
 \widehat{X}(\mathbf{k})= \frac{1}{V_{L}} \int_{V_{L}} X(\mathbf{x}) \exp(i\mathbf{k\cdot x)} \, dV,
\end{equation}
where $X(\mathbf{x}) = W(\mathbf{x}) \sqrt{\rho} {v_{j}}$, for $\mathbf{x} \in \{x,y,z\}$, $W(\mathbf{x})$ is the windowing function, $\rho$ is the density, and $v_{j}$ is the velocity component.
We calculate the spectral turbulent kinetic energy density in each $k$-space shell as:
\begin{equation}
 E(k) = \frac{1}{2} \int_{k{\rm-shell}} \mid \widehat{X}(k) \mid^2\, k^2 \, d\Omega_{k},
\end{equation}
where $k$ is the wavenumber $k = \mid{\mathbf{k}} \mid = (k_{x}^2+k_{y}^2+k_{z}^2)^{1/2}$, and $d\Omega_{k}$ is the solid angle element in the Fourier space. The solid angle is the covered opening measured from a specific point in the $k$-space. We fulfill Parseval's theorem, that is, the sum of the spectral turbulent kinetic energy densities ($e_{\rm kin}$ multiplied by the windowing function $W(x,y,z)$) over $k$-space equals the total energy densities in real-space:
\begin{equation}
 \int_{k_{\rm min}}^{k_{\rm max}} \widehat{e}_{\rm kin} \, dk = \int_{V_{L}} W(x,y,z) e_{\rm kin} \, dV,
\end{equation}
Finally, we compute the normalized turbulent kinetic energy power spectrum as:

\begin{equation}
 \widehat{E}(k)=\Bigg[\int_{k_{\rm min}}^{k_{\rm max}} \widehat{e}_{\rm kin}(k)~dk\Bigg]^{-1}~~\widehat{e}_{\rm kin}(k).
\end{equation}

\begin{figure}[hp!]
    \centering
    \includegraphics[width=0.9\textwidth,height=9.5cm]{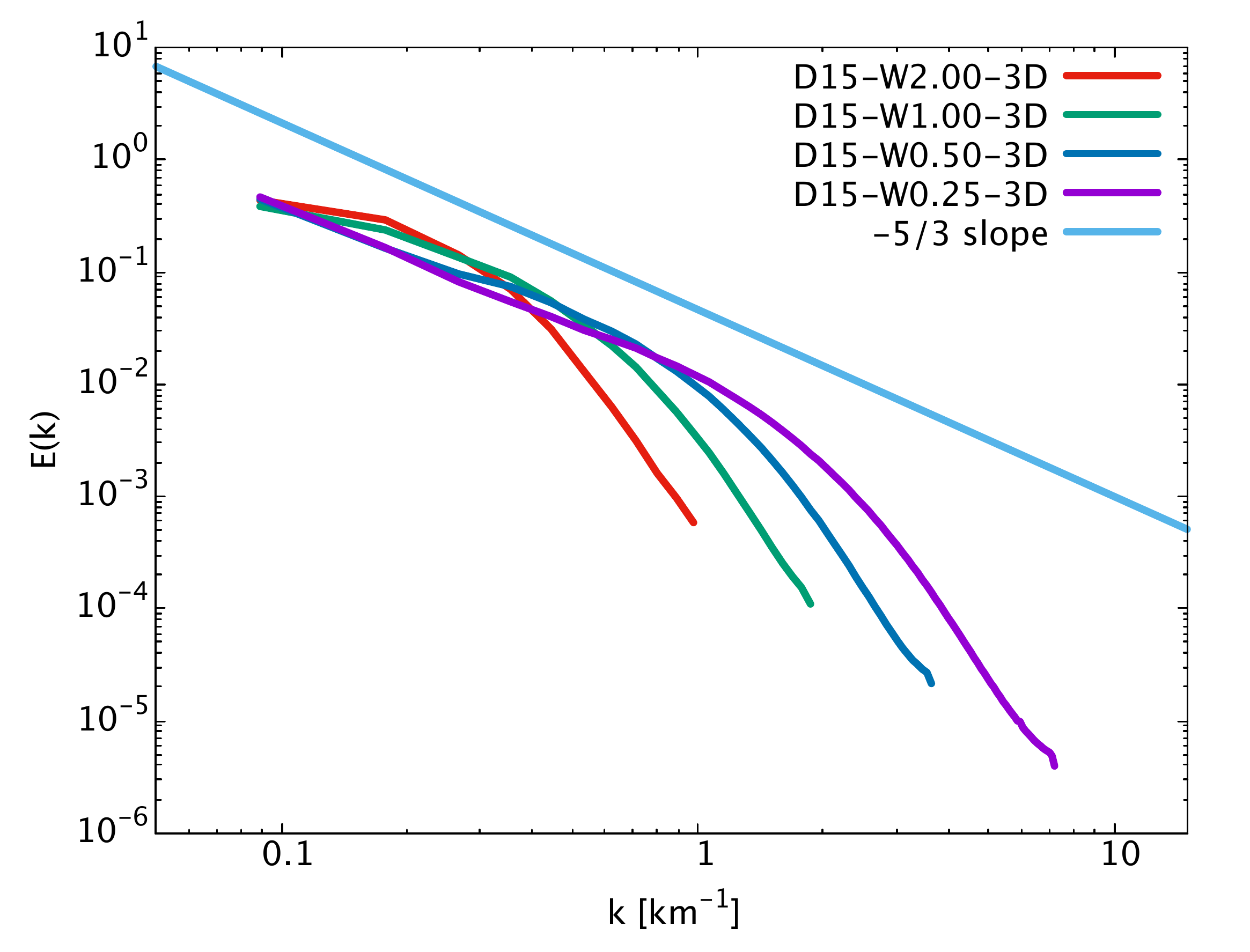}\\
    \includegraphics[width=0.9\textwidth,height=9.5cm]{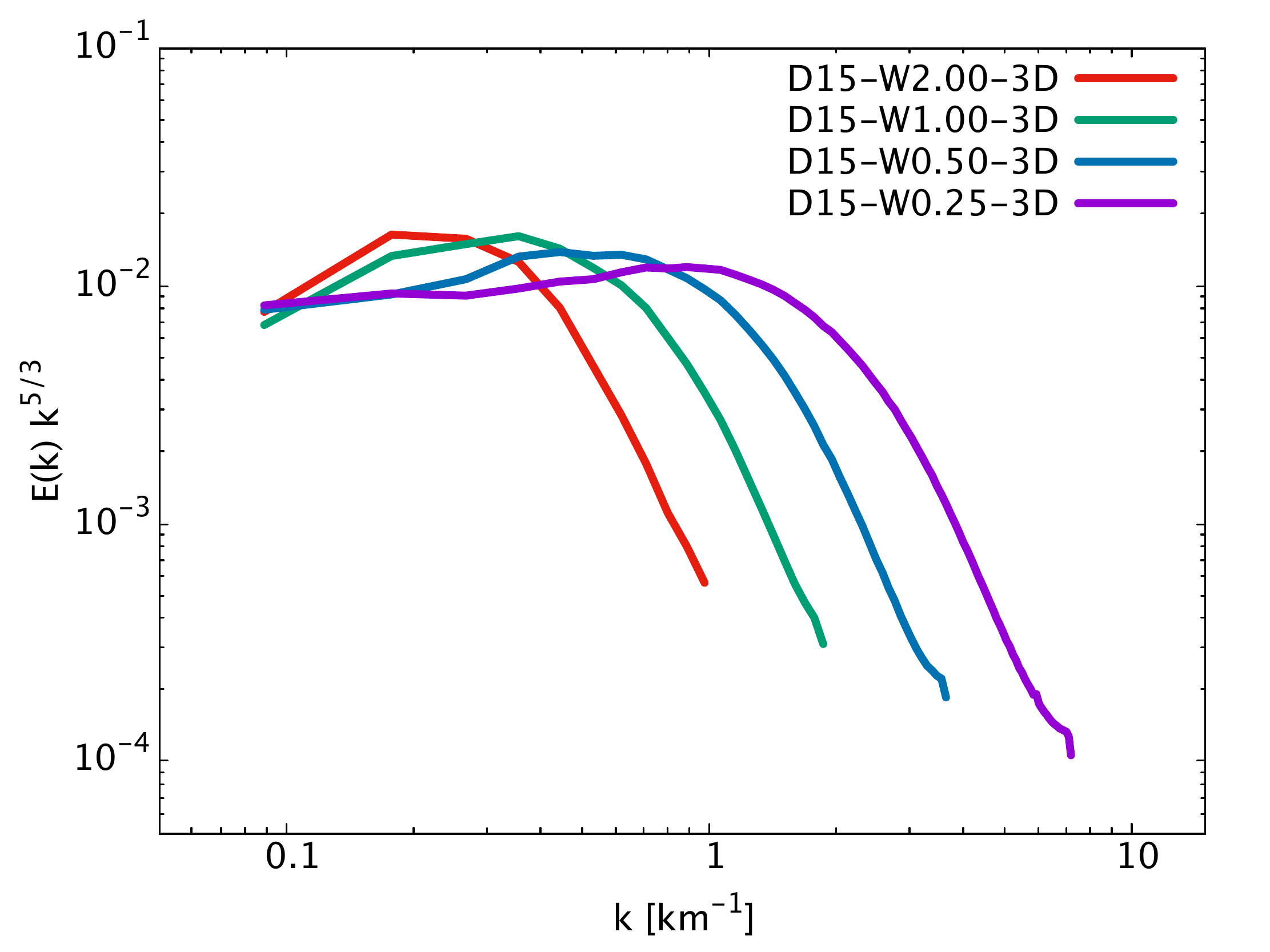}\\
    \caption{Upper panel: turbulent kinetic energy power spectrum of the 3D models. The $k^{-5/3}$ scaling of Kolmogorov's theory is displayed for comparison. Lower panel: compensated kinetic turbulent energy power spectrum ($E(k)\,k^{5/3}$ versus $k$) of the 3D models.}
\label{fig:PSWedge}
\end{figure}

\subsection{Turbulent kinetic energy power spectra of wedge models}

We computed the normalized time-averaged turbulent kinetic energy power spectrum as a function of the wavenumber for the 3D models (see upper panel of Figure \ref{fig:PSWedge}), from t$\sim$150~ms to the end of the simulation: t$\sim$285.0~ms for D15-W2.00-3D, t$\sim$297.0~ms for D15-W1.00-3D, t$\sim$282.2~ms for D15-0.50-3D, and t$\sim$195.0~ms for D15-W0.25-3D.
The slope of the power spectra becomes steeper when the resolution increases, transitioning from shallow slopes of $\sim$-1 (D15-W2.00-3D and D15-W1.00-3D) to slopes of $\sim$-4/3 to -5/3 (D15-W0.50-3D and D15-W0.25-3D). 
Thus, high-resolution models start recovering the $k^{-5/3}$ scaling predicted by Kolmogorov's theory \cite{Kolm41}.
We present the compensated power spectrum, $E(k)~k^{5/3}$, in the lower panel of Figure \ref{fig:PSWedge}. 
It is worth noting the presence of notable peaks, especially at low resolutions, which are the result of an accumulation of energy at large scales ($k\sim 0.2$~km$^{-1}$ for  D15-W2.00-3D and $k\sim0.3$~km$^{-1}$ for  D15-W1.00-3D). The presence of shallow slopes and pronounced peaks at low resolutions are also found in previous studies \cite{DoBuMu13,CoOc14,AbOtRa15,CoOt15,RaCoOt15}.
The peaks may be due to a numerical artifact called the bottleneck effect \cite{ShJa93,FrKuPa08,DoHaYo03}.
Owing to an insufficient resolution, the interaction among eddies of different sizes is impeded, thus preventing the expected energy transfer down to small scales, such that more energy resides at larger scales.
This is consistent with the presence of long-lived, large-scale structures seen in our low-resolution simulations.
High-resolution models instead exhibit short-lived, small scale structures.  
Therefore, interactions among eddies becomes more effective, which in turn, result in a smoother energy transfer from large to small scales.
Consequently, D15-0.50-3D and D15-0.25-3D show attenuated peaks in the compensated power spectrum and slopes more compatible with the predicted Kolmogorov scaling \cite{RaOtAb16}.
Whether the build up of turbulent power at large scales or dissipation at smaller scales following the cascade to smaller scales impacts shock revival will be examined through the contributions to the Reynolds stress in Section~\ref{sec:reynolds}.


\begin{figure}
    \centering
    \includegraphics[width=0.9\textwidth,height=9.5cm]{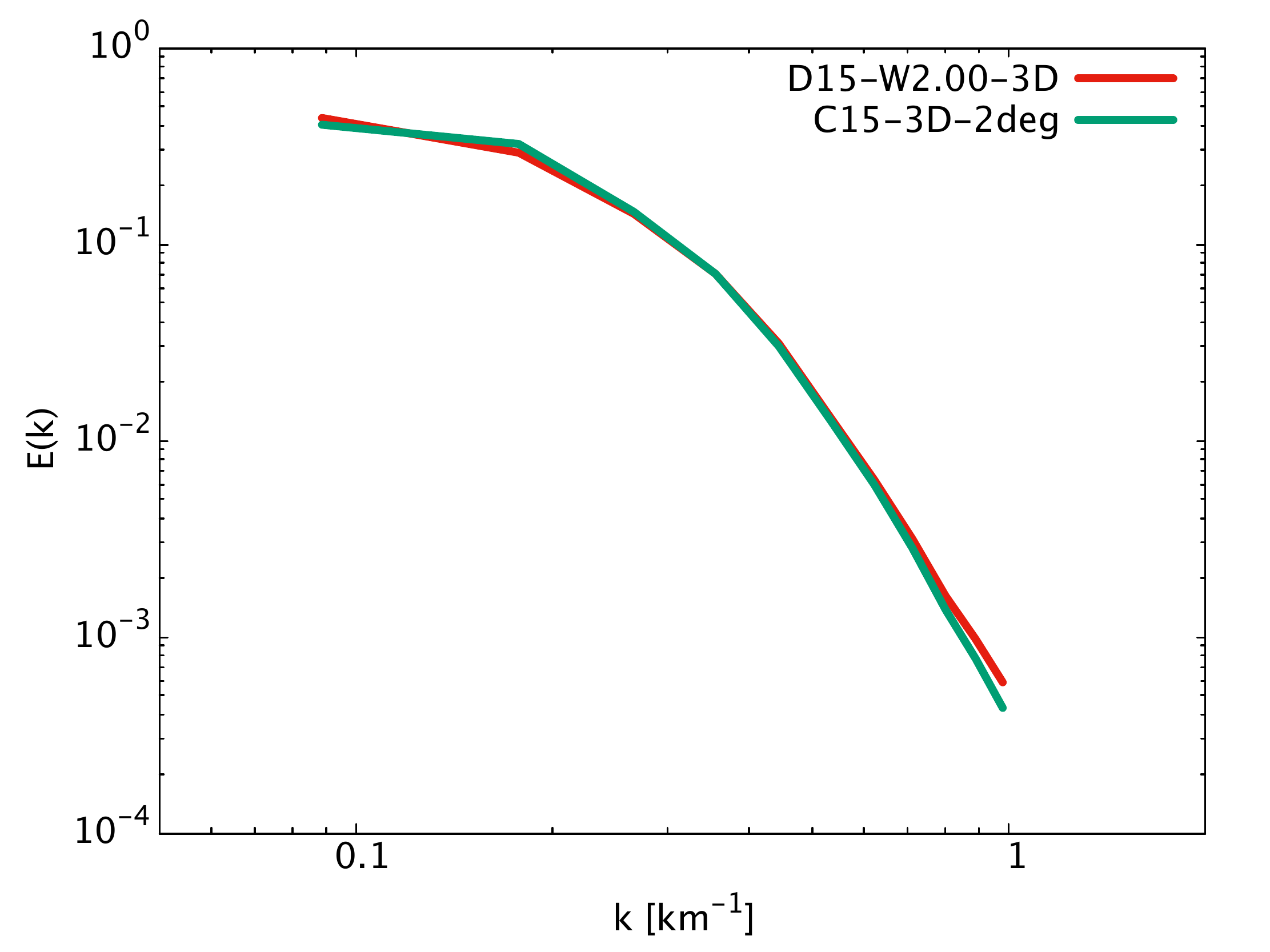}\\
    \includegraphics[width=0.9\textwidth,height=9.5cm]{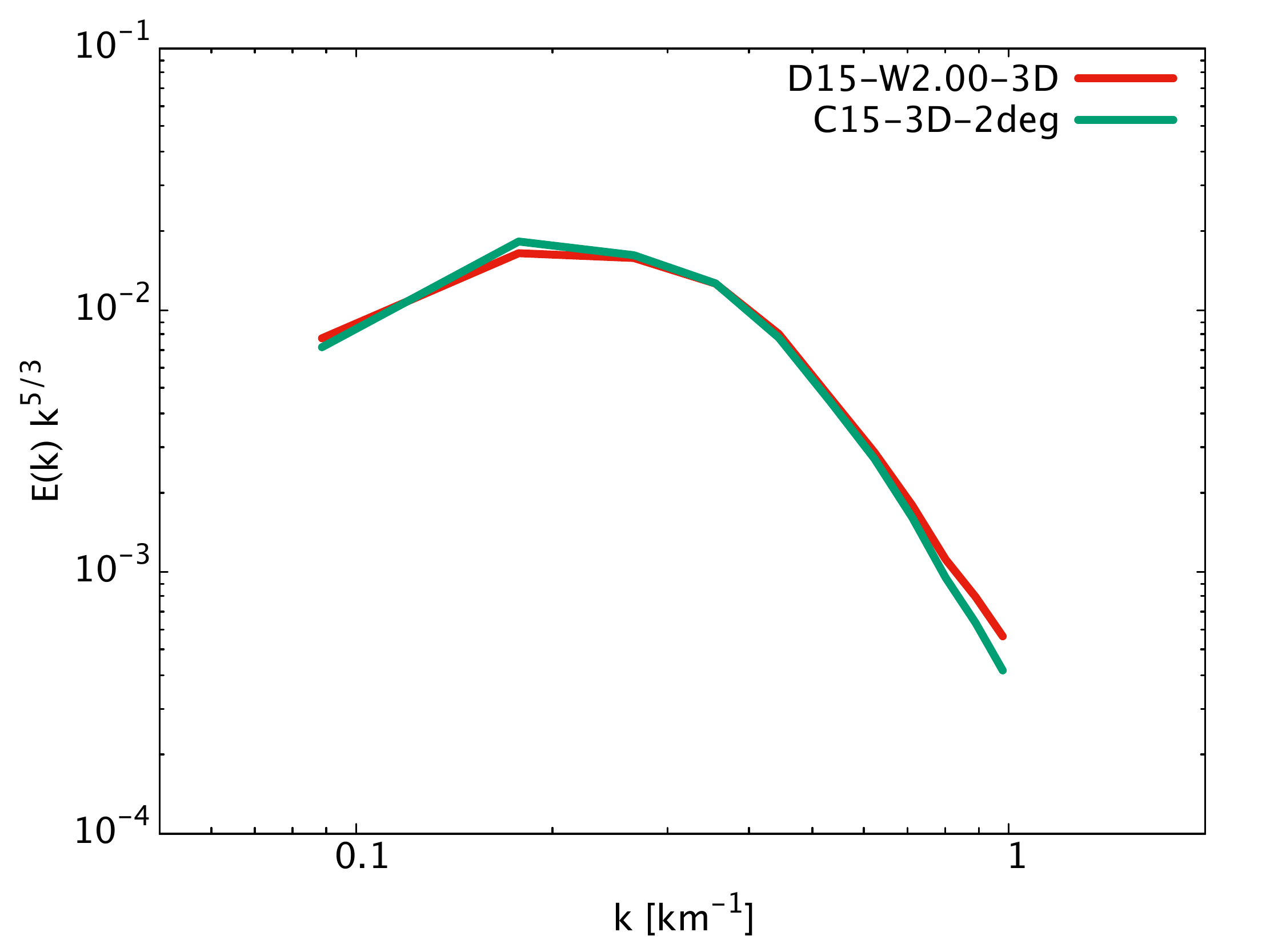}\\
    \caption{Upper panel: comparison of the turbulent kinetic energy power spectrum between D15-W2.00-3D (wedge) and C15-3D-2deg (full) using the same cubic box configuration of 50$\sqrt{2}$$\times$50$\sqrt{2}$$\times$50$\sqrt{2}$~km$^{3}$, placed in the gain region. Lower panel: comparison of the compensated turbulent kinetic energy power spectrum ($E(k)\,k^{5/3}$ versus $k$) between C15-W2.00-3D (wedge) and D15-3D-2deg (full).}
\label{fig:PSWedgeFull}
\end{figure}

\subsection{Turbulent kinetic energy power spectra of full models: Comparison between full and wedge models}
We calculated the turbulent kinetic energy power spectrum of a full 4$\pi$ model with an angular resolution of 2$^\circ$ (C15-3D-2deg in Table 1). This model is a 2$^\circ$ resolution counterpart of the model in \cite{LeBrHi15}. Initial conditions were also taken from the 15~\msun\ pre-supernova progenitor of \cite{WoHe07}, and the simulation was computed from a 1D simulation at 1.3~ms after bounce by applying a 0.1\% density perturbation over 
radii 10--30~km, mimicking perturbations seen in simulations evolved through bounce in 2D. The angular computational domain spans the full sphere $\theta\in[0,\pi]$ and $\phi\in[0,2\pi]$, respectively. Model C15-3D-2deg was initialized with a 180-zone $\phi$-grid ($\delta\phi$=2$^\circ$) and a 90-zone $\theta$-grid ($\delta\theta$=2$^\circ$).

In the upper and lower panels of Figure \ref{fig:PSWedgeFull}, we present a comparison of the turbulent kinetic energy power spectrum and compensated power spectrum, respectively, between 2$^\circ$ resolution wedge (D15-W2.00-3D) and 4$\pi$ (C15-3D-2deg) models \footnote[3]{The time-averaged turbulent kinetic energy power spectrum for C15-3D-2deg is computed over the time extending from t=150~ms to t=400~ms.}. 
We adopted the same cubic box of (50$\sqrt{2}$~km)$^{3}$ placed in the gain region for both models.
 Figure~\ref{fig:PSWedgeFull} show that the computational domain ($90^{\circ}$ wedge model versus full $4\pi$ model) has a minor impact on the power spectrum at the scales captured within the analysis box.
Therefore, a reduced computational domain (i.e., the 90$^\circ$ computational domain utilized in the wedge model) does not introduce numerical artifacts and is a robust configuration to perform a numerical resolution study of turbulence during the initial explosion phase of CCSNe.   

\section{Reynolds Stress}
\label{sec:reynolds}
Recent studies \cite{MuDoBu13,CoOt15,RaOtAb16} suggest that turbulent convection contributes an additional source of pressure (stress) that aids in counteracting the ram pressure of the infalling material, such that the stalled shock is revived. According to \cite{MuDoBu13}, the Reynolds decomposition at the shock jump provides a means to study this turbulent pressure component. The Reynolds stress measures the mean pressure induced by turbulent fluctuations and its radial component is calculated as $R_{rr}=\langle \rho v'_{r} v'_{r}\rangle/\langle \rho \rangle$, where $v'_{r}(r,\theta,\phi)=\langle v_{r}\rangle(r)-v_{r}(r,\theta,\phi)$ \cite{BrLeHi16} and $\langle \cdot \rangle$ is the angle, or shell, averaging operator. 

\begin{figure}
    \centering
    \includegraphics[width=0.9\textwidth,height=8cm]{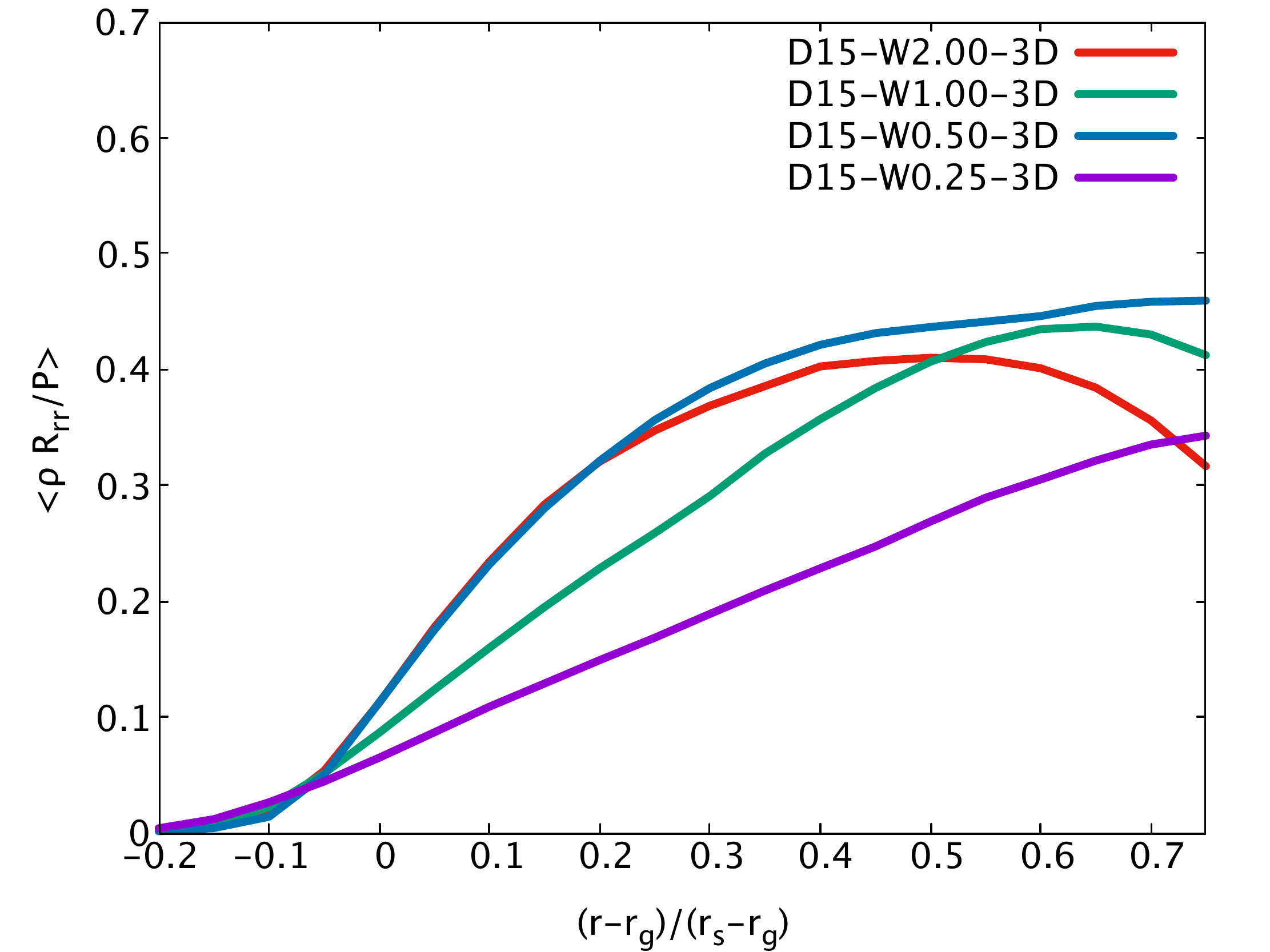}
    \caption{Time-averaged ratio of the radial Reynolds stress and the thermal pressure for D15-W2.00-3D, D15-W1.00-3D, D15-0.50-3D, and D15-W0.25-3D, as a function of $(r-r_{g})$/$(r_{s}-r_{g})$, where $r_{g}$ is the gain radius and $r_{s}$ is the shock radius.}
\label{fig:Reynolds}
\end{figure}

In Figure \ref{fig:Reynolds}, we present the angle-averaged ratio between the radial Reynolds stress and the thermal pressure, $\langle \rho R_{rr}/P\rangle$, for the 3D models, from t=150.4~ms to t=282.2~ms, except for D15-W0.25-3D, as it was computed until t$\sim$195~ms.  
The turbulent pressure represents $\sim$40--50\% of the total pressure for D15-W1.00-3D and D15-W0.50-3D, while the contribution is smaller for D15-W2.00-3D. Therefore, resolving turbulence may be crucial for the development of the explosion, as it provides an extra pressure that results in a shock revival. The values of the contribution of the turbulent pressure prior to the shock revival are consistent with those reported in previous works ($\sim$50\% in \cite{CoOt15}, and $\sim$30--40\% in \cite{RaOtAb16}).

\section{Discussion}

Previous parameterized multidimensional studies reveal that turbulence can play a role in reviving the shock, as it provides an additional pressure ($\sim$30--40\% of the thermal pressure) that aids in pushing the shock outwards \cite{MuDoBu13,CoOt15,RaOtAb16}.
We have analyzed the effect of numerical resolution on turbulence in CCSN simulations.  
We employed our neutrino radiation hydrodynamics code \Chimera\ \cite{BrBlHi18} to perform four 3D simulations with a reduced angular extent, using a 15~\msun\ progenitor model at different angular resolutions (2$^\circ$, 1$^\circ$, 1/2$^\circ$, 1/4$^\circ$). The turbulent kinetic energy power spectra for low resolution models exhibit shallow slopes that progressively steepen as the resolution increases. The energy cascade to small scales appears to be influenced by the bottleneck effect. It truncates or, as is the case for our lowest resolution model, eliminates any inertial range, and turbulent energy accumulates on large scales. 
Previous studies using parameterized models have argued that the bottleneck effect may lead to spurious or premature explosions \cite{DoBuMu13,CoOc14,AbOtRa15,CoOt15,RaOtAb16} at low resolution.  Our models lack conclusive evidence of this, given no clear trends with resolution in the computed shock progression, turbulent kinetic energy, or Reynolds stresses. However, in agreement with the aforementioned studies, the bottleneck effect softens at higher resolutions, and our models start showing an extended region compatible with the predicted Kolmogorov scaling, with a spectral slope of $E(k)\propto k^{-5/3}$ (especially for model D15-W0.25-3D).  


The reduced angular extent (90$^\circ$ computational domain) deployed to achieve high resolution may impose some restrictions on the study of turbulence characteristics by introducing boundary effects and excluding large scale dynamics of a full model.  
However, a comparison of the turbulent kinetic energy power spectrum between a full and a wedge model confirms that the limited domain does not introduce noticeable artifacts, and that the wedge format is an adequate configuration to capture convective features and to study the character of turbulence in CCSNe.  

We have quantified the relative contribution of the radial Reynolds stress to the thermal pressure in our 3D models. Turbulent stresses adds $\sim$40\% to the total pressure (values in the range of 30--50\% are reported in previous studies), reaching $\sim$45\% for  D15-W0.50-3D. The turbulent pressure contributes throughout the gain region, particularly near the shock. This is consistent with previous results \cite{CoOt15,RaOtAb16}, and suggests that turbulent pressure may be an important factor in the shock revival. However, our simulations are not evolved for a sufficient duration to determine the impact of resolution and turbulence on explodability.  

By performing an examination of turbulence in CCSNe using the full complement of physics used in sophisticated explosion models, excepting for the use of the reduced angular domain, we have demonstrated the viability of parameterized models for high-resolution studies of CCSN turbulence.  Firstly, we demonstrated that the full-physics models, like their more parameterized counterparts, show the bottleneck effect at lower resolutions with the development of an inertial range and $E(k) \sim k^{-5/3}$ Kolmogorov scaling at higher resolution. Secondly, the comparison of turbulent properties of a wedge and $4\pi$ model of identical resolution demonstrates that the properties of turbulence in the analysis box are captured in the wedge model showing that the restricted domain can realistically sample the turbulent behavior found in full models at smaller scales.  Together, these show the continued usefulness of high-resolution, parameterized models in the study of the properties of turbulence in CCSNe.  The larger question of how turbulence impacts the development of explosions and how truncating the turbulent cascade at larger scales impacts lower-resolution, full CCSN models is not clear.  It is known that large-scale modes are present in simulated CCSNe explosions that can't be captured in the restricted domain simulations and that parameterized models can be tuned to drive explosions without relying on natural feedbacks in the simulation.  However, studies of turbulent flows in limited-domain simulations and parameterized simulations at higher resolutions than can be afforded for full (physics, domain, and duration) simulations are important. These studies can aid in understanding the role played by turbulence in the limited examinations of resolution in full simulations \cite{MuDoBu13,CoOt15} that are at the lower end of the resolutions studied in this and similar turbulent CCSN studies \cite{RaOtAb16,NaBuRa19,MeJa19}.  

\section*{Acknowledgements}

This research was supported by the U.S. Department of Energy Offices of Nuclear Physics and Advanced Scientific Computing Research and
an award of computer time provided by the Innovative and Novel Computational Impact on Theory and Experiment (INCITE) program at the Oak Ridge Leadership Computing Facility (OLCF), which is a DOE Office of Science User Facility supported  under contracts DE-AC05-00OR22725.
This research is part of the Blue Waters sustained-petascale computing project, which is supported by the National Science Foundation (awards OCI-0725070 and ACI-1238993) and the state of Illinois.
Blue Waters is a joint effort of the University of Illinois at Urbana-Champaign and its National Center for Supercomputing Applications.
This work is also part of the ``Core-collapse Supernovae Through Cosmic Time'' Petascale Computational Resource (PRAC) allocation support by the National Science Foundation (award ACI-144005).

%

%
%

\end{document}